\documentclass[12pt,a4paper,twoside,english]{article}
\usepackage[T1]{fontenc}
\usepackage[latin1]{inputenc}
\usepackage{amssymb}

\makeatletter

\usepackage{babel}
\makeatother

\begin{document}

\title{Electromagnetic eigenmodes in matter. van der Waals-London and Casimir
forces}

\author{M. Apostol and G. Vaman \\
Department of Theoretical Physics, Institute of Atomic Physics, \\
Magurele-Bucharest MG-6, POBox MG-35, Romania \\
email: apoma@theory.nipne.ro}

\date{{}}

\maketitle
\begin{abstract}
We derive van der Waals-London and Casimir forces by calculating the
eigenmodes of the electromagnetic field interacting with two semi-infinite
bodies (two halves of space) with parallel surfaces separated by distance
$d$. We adopt simple models for metals and dielectrics, well-known
in the elementary theory of dispersion. In the non-retarded (Coulomb)
limit we get a $d^{-3}$-force (van der Waals-London force), arising
from the zero-point energy (vacuum fluctuations) of the surface plasmon
modes. When retardation is included we obtain a $d^{-4}$-(Casimir)
force, arising from the zero-point energy of the surface plasmon-polariton
modes (evanescent modes) for metals, and from propagating (polaritonic)
modes for identical dielectrics. The same Casimir force is also obtained
for \char`\"{}fixed surfaces\char`\"{} boundary conditions, irrespective
of the pair of bodies. The approach is based on the equation of motion
of the polarization and the electromagnetic potentials, which lead
to coupled integral equations. These equations are solved, and their
relevant eigenfrequencies branches are identified. 
\end{abstract}


\section{Introduction}

The Casimir force was originally derived by estimating the zero-point
energy (vacuum fluctuations) of the electromagnetic field comprised
in-between two ideal, perfectly reflecting, semi-infinite metals (two
halves of space) separated by distance $d$.\cite{key-1} As it is
well-known, it goes like $d^{-4}$ for distances greater than the
characteristic electromagnetic wavelengths of the bodies (plasmon
\char`\"{}wavelengths\char`\"{}). Further on, the calculations have
been cast in a different form, by resorting to the fluctuations theory,\cite{key-2,key-3}
and a $d^{-3}$-force has been obtained for the non-retarded (Coulomb)
interaction, which corresponds to the van der Waals-London force.
The matter polarization is usually represented in this case by a dielectric
function. Recently, there is a renewed interest in this subject, motivated,
on one hand, by the role played by plasmons, polaritons and other
surface effects arising from the interaction between the electromagnetic
field and matter and, on the other hand, by the querries related to
the applicability of a dielectric function for discontinuous bodies.\cite{key-4}-\cite{key-20}
We report here on a different investigation of these forces, based
on the calculation of the eigenfrequencies of the electromagnetic
field interacting with matter. 

We assume a simple model of matter, consisting of mobile particles
with charge $-e$ and mass $m$, moving in a rigid neutralizing background,
and subjected to certain forces. Such a model is reminiscent of the
well-known jellium model of electron plasma, though it is generalized
here to some extent. In the presence of the electromagnetic field
matter polarizes. We leave aside the magnetization (we consider only
non-magnetic matter) and relativistic effects. We represent the small
disturbance in the density of the mobile charges as $\delta n=-ndiv\mathbf{u}$,
where $n$ is the (constant) concentration of the charges and $\mathbf{u}$
is a displacement field in the positions of these charges. The charge
disturbance is therefore $\rho=endiv\mathbf{u}$. This representation
is valid for $\mathbf{K}\mathbf{u}(\mathbf{K})\ll1$, where $\mathbf{K}$
is the wavevector and $\mathbf{u}(\mathbf{K})$ is the Fourier transform
of the displacement field. 

For homogeneous and isotropic matter the displacement field obeys
an equation of motion which can be taken of the form \begin{equation}
m\ddot{\mathbf{u}}=-e\mathbf{E}-e\mathbf{E}_{0}-m\omega_{0}^{2}\mathbf{u}-m\gamma\dot{\mathbf{u}}\,\,\,,\label{1}\end{equation}
 where \textbf{$\mathbf{E}$ }is the (internal) electric field, $\mathbf{E}_{0}$
is an external electric field, $\omega_{0}$ is a frequency parameter
corresponding to an elastic force and $\gamma$ is a dissipation parameter.
Making use of the temporal Fourier transform we get \begin{equation}
\mathbf{u}(\omega)=\frac{e}{m}\frac{1}{\omega^{2}-\omega_{0}^{2}+i\omega\gamma}(\mathbf{E}+\mathbf{E}_{0})\label{2}\end{equation}
(where we dropped out the argument $\omega$ of the electric fields).
On the other hand, from Maxwell's equation $div\mathbf{E}=4\pi endiv\mathbf{u}$,
we get the (internal) electric field $\mathbf{E}=4\pi ne\mathbf{u}$
(equal to $-4\pi\mathbf{P}$, where $\mathbf{P}$ is the polarization).
Making use of equation (\ref{2}) we get the dielectric function \begin{equation}
\varepsilon(\omega)=1-\frac{\omega_{p}^{2}}{\omega^{2}-\omega_{0}^{2}+i\omega\gamma}\label{3}\end{equation}
from its definition $\mathbf{E}_{0}=\varepsilon(\mathbf{E}+\mathbf{E}_{0})$,
where $\omega_{p}$, given by $\omega_{p}^{2}=4\pi ne^{2}/m$, is
the plasma frequency. The dielectric function given by equation (\ref{3})
is well known in the elementary theory of dispersion.\cite{key-21}
It proves to be a fairly adequate representation for matter polarization
in various bodies. We can view $\omega_{p},$$\omega_{0}$ and $\gamma$
as free parameters, thus being able to simulate various models of
matter. For $\omega_{0}=\gamma=0$ we get the well-known dielectric
function of an ideal plasma; if $\omega_{0}=0$ we have the dielectric
function of the optical properties of simple metals for $\omega\gg\gamma$
(Drude model), and the dielectric function corresponding to the static
(or quasi-static) currents in metals for $\omega\ll\gamma$; for $\omega_{0}\gg\omega_{p}$
we have a dielectric function of dielectrics with loss; and so on. 

In addition, making use of equation (\ref{2}), we can compute also
the electric conductivity $\sigma$, from its definition $\mathbf{j}=\sigma(\mathbf{E}+\mathbf{E}_{0})$,
where $\mathbf{j}=-en\dot{\mathbf{u}}$ is the current density. We
get the well-known conductivity \begin{equation}
\sigma(\omega)=\frac{\omega_{p}^{2}}{4\pi}\frac{i\omega}{\omega^{2}-\omega_{0}^{2}+i\omega\gamma}\,\,\,,\label{4}\end{equation}
 whence, for instance, the static conductivity for metals $\sigma=\omega_{p}^{2}/4\pi\gamma$;
parameter $\gamma$ can be viewed as the reciprocal of a damping time
$\tau$ (or relaxation time, or lifetime), $\gamma=1/\tau$, and we
get the well-known static conductivity $\sigma=ne^{2}\tau/m\gamma$.

Therefore, the equation of motion (\ref{1}) turns out to be an adequate
starting point for representing the matter polarization. However,
we must note that for dielectrics, which may imply oscillations in
localized atoms (in our model through the frequency $\omega_{0}$),
the classical dynamics assumed here turns out to be inadequate in
the retarded regime, and a quantum treatment is then required. 

In the non-retarded limit the electric field $\mathbf{E}$ in equation
(\ref{1}) is given by the Coulomb law, \emph{i.e.} $\mathbf{E}=-grad\Phi$,
where $\Phi$ is the static Coulomb potential arising in matter. The
latter depends on the charge disturbance $\rho=-e\delta n$, therefore
on $\mathbf{u}$. Then, it is easy to see that the equation of motion
(\ref{1}) leads to an integral equation for the displacement field
$\mathbf{u}$. Its eigenvalues give the plasmon modes. For retarded
interaction, the electric field $\mathbf{E}$ in equation (\ref{1})
is given by the vector potential $\mathbf{A}$ and the scalar potential
$\Phi$ trough $\mathbf{E}=-\frac{1}{c}\frac{\partial\mathbf{A}}{\partial t}-grad\Phi$.
Making use of the radiation (Kirchhoff) formulae, these potentials
can be expressed as integrals containing the displacement field $\mathbf{u}$
(through the charge and current densities), and we get again an integral
equation for $\mathbf{u}$. Its eigenvalues give polariton-like modes.
The use of integral equations in treating the electromagnetic field
interacting with matter was previously indicated in connection with
the so-called Ewald-Oseen extinction theorem.\cite{key-22} We have
applied this approach to a semi-infinite (half-space) body, as well
as to a slab of finite thickness.\cite{key-23} In this case, beside
the bulk displacement field, there appears a surface displacement
field also, and the integral equations couple these degrees of freedom.
We have solved these coupled integral equations and computed bulk
and surface plasmons and polaritons, dielectric response, reflected,
refracted and transmitted fields, and derived generalized Fresnel
relations. We employ the same procedure here for two semi-infinite
bodies (two halves of the space) separated by distance $d$, in order
to get the electromagnetic eigenfrequencies and to derive van der
Waals-London and Casimir forces. We do it in two steps: first, for
static Coulomb (non-retarded) interaction (valid for wavelengths much
longer than the characteristic size of the bodies) and, second, for
retarded interaction.

\section{Surface plasmons. van der Waals-London forces}

We consider two semi-infinite bodies (two halves of space) with parallel
surfaces in the $(x,y)$-plane, separated by distance $d$. The bodies
occupy the regions $z<-d/2$ and, respectively, $z>d/2$. We take
two displacement fields $\mathbf{u}_{1,2}$, giving rise to two charge
disturbances $\delta n_{1,2}=-n_{1,2}div\mathbf{u}_{1,2}$. We consider
first the equation of motion for an ideal plasma. In general, we leave
aside the dissipation (parameter $\gamma$ in equation (\ref{1})),
which is irrelevant for our discussion. The equation of motion reads
\begin{equation}
m\ddot{\mathbf{u}}_{1}=grad\int d\mathbf{R}^{'}U(\left|\mathbf{R}-\mathbf{R}^{'}\right|)\left[n_{1}div\mathbf{u}_{1}(\mathbf{R}^{'})+n_{2}div\mathbf{u}_{2}(\mathbf{R}^{'})\right]\,\,\,,\label{5}\end{equation}
and a similar equation for $\mathbf{u}_{2}$, which can be obtained
from equation (\ref{5}) by interchanging the labels $1$ and $2$
($1\longleftrightarrow2$); $U(R)=e^{2}/R$ in equation (\ref{5})
is the Coulomb interaction. Since we are interested in the eigenmodes,
we leave aside the external field $\mathbf{E}_{0}$. We use $\mathbf{R}=(\mathbf{r},z)$
for the position vector $\mathbf{R}$, where $\mathbf{r}=(x,y),$
and the representation \begin{equation}
\mathbf{u}_{1,2}=(\mathbf{v}_{1,2},w_{1,2})\theta(\pm z-d/2)\label{6}\end{equation}
 for the displacement fields, where $\theta(z)=1$ for $z>0$ and
$\theta(z)=0$ for $z<0$ is the step function; the $\pm$ sign is
associated with labels $1$ and $2$, respectively. The divergence
in equation (\ref{5}) can now be written as \begin{equation}
div\mathbf{u}_{1,2}=\left(div\mathbf{v}_{1,2}+\frac{\partial w_{1,2}}{\partial z}\right)\theta(\pm z-d/2)+w_{1,2}(\pm d/2)\delta(\pm z-d/2)\,\,\,,\label{7}\end{equation}
where $w_{1,2}(\pm d/2)$ means $w_{1,2}(\mathbf{r},z=\pm d/2)$.
We notice in equation (\ref{7}) the (de)polarization charge arising
at the surfaces $z=\pm d/2$. We employ Fourier representations of
the form \begin{equation}
\mathbf{v}_{1,2}(\mathbf{r},z;t)=\sum_{\mathbf{k}}\int d\omega\mathbf{v}_{1,2}(\mathbf{k},z;\omega)e^{i\mathbf{kr}}e^{-i\omega t}\label{8}\end{equation}
 and similar ones for $w_{1,2}$, and use the Fourier transform \begin{equation}
\frac{1}{\sqrt{r^{2}+z^{2}}}=\sum_{\mathbf{k}}\frac{2\pi}{k}e^{-k\left|z\right|}e^{i\mathbf{kr}}\label{9}\end{equation}
 for the Coulomb potential. Then, we notice that equation (\ref{5})
implies that $\mathbf{v}_{1,2}$ are parallel with the wavevector
$\mathbf{k}$ (in-plane \char`\"{}longitudinal\char`\"{} modes), and
$i\mathbf{k}w_{1,2}=\frac{\partial\mathbf{v}_{1,2}}{\partial z}$.
We use this latter relation to eliminate $w_{1,2}$ from the equations
of motion. In addition, we introduce the notation $v_{1,2}=\mathbf{k}\mathbf{v}_{1,2}/k$.
Then, it is easy to see that equation (\ref{5}) yields two coupled
integral equations \begin{equation}
\begin{array}{c}
\omega^{2}v_{1}=\frac{\omega_{1}^{2}k}{2}\int_{d/2}^{\infty}dz^{'}e^{-k\left|z-z'\right|}v_{1}+\frac{\omega_{1}^{2}}{2k}\int_{d/2}^{\infty}dz^{'}\frac{\partial}{\partial z^{'}}e^{-k\left|z-z'\right|}\frac{\partial v_{1}}{\partial z^{'}}+\\
\\+\frac{\omega_{2}^{2}k}{2}\int_{-\infty}^{-d/2}dz^{'}e^{-k(z-z')}v_{2}+\frac{\omega_{2}^{2}}{2}\int_{-\infty}^{-d/2}dz^{'}e^{-k(z-z')}\frac{\partial v_{2}}{\partial z^{'}}\,\,,\,\, z>d/2\,\,\,,\\
\\\omega^{2}v_{2}=\frac{\omega_{1}^{2}k}{2}\int_{d/2}^{\infty}dz^{'}e^{k(z-z')}v_{1}-\frac{\omega_{1}^{2}}{2}\int_{d/2}^{\infty}dz^{'}e^{k(z-z')}\frac{\partial v_{1}}{\partial z^{'}}+\\
\\+\frac{\omega_{2}^{2}k}{2}\int_{-\infty}^{-d/2}dz^{'}e^{-k\left|z-z'\right|}v_{2}+\frac{\omega_{2}^{2}}{2k}\int_{-\infty}^{-d/2}dz^{'}\frac{\partial}{\partial z^{'}}e^{-k\left|z-z'\right|}\frac{\partial v_{2}}{\partial z^{'}}\,\,,\,\, z<-d/2\,\,\,,\end{array}\label{10}\end{equation}
 where $\omega_{1,2}^{2}=4\pi n_{1,2}e^{2}/m$ and we dropped out
the arguments $\omega,\mathbf{k}$. Integrating by parts in equations
(\ref{10}) we obtain a system of two algebraic equations \begin{equation}
\begin{array}{c}
\left(\omega^{2}-\omega_{1}^{2}\right)v_{1}=-\frac{1}{2}e^{-kz}\left[\omega_{1}^{2}e^{kd/2}v_{1}(d/2)-\omega_{2}^{2}e^{-kd/2}v_{2}(-d/2)\right]\,\,,\,\, z>d/2\,\,\,,\\
\\\left(\omega^{2}-\omega_{2}^{2}\right)v_{2}=\frac{1}{2}e^{kz}\left[\omega_{1}^{2}e^{-kd/2}v_{1}(d/2)-\omega_{2}^{2}e^{kd/2}v_{2}(-d/2)\right]\,\,,\,\, z<-d/2\,\,.\end{array}\label{11}\end{equation}
 We can see that in this non-retarded limit the two bodies are coupled
only through their surfaces. 

For $v_{1}(d/2)=v_{2}(-d/2)=0$ in equations (\ref{11}) we get the
bulk plasmons $\omega=\omega_{1,2}$. Making $z=\pm d/2$ in equations
(\ref{11}) we get the system of equations for the surface modes.
The corresponding dispersion equation is given by \begin{equation}
\left(\omega^{2}-\frac{1}{2}\omega_{1}^{2}\right)\left(\omega^{2}-\frac{1}{2}\omega_{2}^{2}\right)-\frac{1}{4}\omega_{1}^{2}\omega_{2}^{2}e^{-2kd}=0\,\,.\label{12}\end{equation}
For $d=0$ we obtain the surface plasmon of a metallic interface given
by $\omega^{2}=\frac{1}{2}\left(\omega_{1}^{2}+\omega_{2}^{2}\right)$,
while for $d\rightarrow\infty$ we get the surface plasmons $\omega=\omega_{1,2}/\sqrt{2}$
for free (uncoupled) surfaces. If the body labelled by $2$ for instance
is a dielectric, then $\omega^{2}$ in the second equation (\ref{11})
is replaced by $\omega^{2}-\omega_{0}^{2}$. In the limit $\omega_{0}\gg\omega_{2}$
and for $d=0$ we get the surface plasmon $\omega=\omega_{1}/\sqrt{1+\varepsilon_{2}}$,
corresponding to a dielectric-metal interface, where $\varepsilon_{2}=1+\omega_{2}^{2}/\omega_{0}^{2}$.
For two identical metals $\omega_{1}=\omega_{2}=\omega_{p}$ we get
the surface plasmons given by\begin{equation}
\omega^{2}=\frac{1}{2}\omega_{p}^{2}\left(1\pm e^{-kd}\right)\,\,.\label{13}\end{equation}
 They are identical with the surface plasmons of a plasma slab of
thickness $d$. These are well-known results.\cite{key-24}-\cite{key-31}

Let us label by $\alpha$ all the eigenvalues $\Omega_{\alpha}$ of
the system of equations (\ref{11}). We compute the force acting between
the two bodies by \begin{equation}
F=\frac{\partial}{\partial d}\sum_{\alpha}\frac{1}{2}\hbar\Omega_{\alpha}\,\,\,,\label{14}\end{equation}
where we recognize the zero-point energy of harmonic oscillators.
Although it can be included straightforwardly, it is easy to see that
the temperature plays no significant role, so we may neglect the temperature
effects, as usually. We may also leave aside the bulk plasmons, since
they do not depend on the distance $d$. We are left with the two
surface modes $\Omega_{1,2}$ given by equation (\ref{12}), labeled
by wavevector $\mathbf{k}$. We can see that these eigenvalues are
function of $kd$, so the force depends on distance $d$ as $F\sim1/d^{3}$.
As it is well-known, such a force between two bodies implies an inter-atomic
interaction $\sim1/R^{6}$ , where $R$ is the distance between two
atoms. This is the well-known van der Waals-London interaction.\cite{key-32} 

We compute here the force $F$ for the eigenvalues given by equation
(\ref{13}), \emph{i.e.} for two identical plasmas (metals). Equation
(\ref{14}) gives a force \begin{equation}
F=\frac{\hbar\omega_{p}}{8\pi\sqrt{2}d^{3}}\int_{0}^{\infty}dx\cdot x^{2}e^{-x}\left(\frac{1}{\sqrt{1-e^{-x}}}-\frac{1}{\sqrt{1+e^{-x}}}\right)\label{15}\end{equation}
per unit area. The integral in equation (\ref{15}) is $\simeq4$,
so we get $F\simeq\hbar\omega_{p}/2\pi\sqrt{2}d^{3}$. In like manner
we can compute the force between two (identical) dielectrics, by replacing
$\omega^{2}$ in equation (\ref{13}) by $\omega^{2}-\omega_{0}^{2}$
and taking the limit $\omega_{0}\gg\omega_{p}$. The result is a much
weaker force $F=\hbar\omega_{p}^{4}/128\omega_{0}^{3}d^{3}$. It can
also be written as $F=\hbar\omega_{0}(\varepsilon-1)^{2}/128d^{3}$,
where $\varepsilon\simeq1+\omega_{p}^{2}/\omega_{0}^{2}$ is the (static)
dielectric function in the limit $\omega\ll\omega_{0}$. The same
result is obtained by making use of the formulae given in Ref. \cite{key-32}
for non-retarded interaction within the framework of the fluctuations
theory (equation 82.3 p. 343 in Ref. \cite{key-32}). Making use of
the eigenvalues given by the roots of the dispersion equation (\ref{12}),
we can compute in the same manner the force acting between two distinct
bodies. For instance, we can consider a dielectric-metal pair and
get straightforwardly the force $F=\hbar\omega_{1}\omega_{2}^{2}/32\pi\sqrt{2}\omega_{0}^{2}d^{3}$,
where $\omega_{1}$ belongs to the metal and $\omega_{2},\,\omega_{0}$
represent the dielectric.

\section{Surface plasmon-polariton modes. Casimir force}

We pass now to the retarded interaction. The electric field in equation
(\ref{1}) is given by $E=-\frac{1}{c}\frac{\partial\mathbf{A}}{\partial t}-grad\Phi$,
where $\mathbf{A}$ is the vector potential and $\Phi$ is the scalar
potential. These potentials are given by \begin{equation}
\mathbf{A}(\mathbf{r},z;t)=\frac{1}{c}\int d\mathbf{r}'\int dz'\frac{\mathbf{j}(\mathbf{r}',z';t-R/c)}{R}\label{16}\end{equation}
 and \begin{equation}
\Phi(\mathbf{r},z;t)=\int d\mathbf{r}'\int dz'\frac{\rho(\mathbf{r}',z';t-R/c)}{R}\,\,\,,\label{17}\end{equation}
where \begin{equation}
\mathbf{j}=-en_{1}(\dot{\mathbf{v}}_{1},\dot{w}_{1})\theta(z-d/2)-en_{2}(\dot{\mathbf{v}}_{2},\dot{w}_{2})\theta(-z-d/2)\label{18}\end{equation}
 is the current density, \begin{equation}
\begin{array}{c}
\rho=en_{1}\left(div\mathbf{v}_{1}+\frac{\partial w_{1}}{\partial z}\right)\theta(z-d/2)+w_{1}(d/2)\delta(z-d/2)+\\
\\+en_{2}\left(div\mathbf{v}_{2}+\frac{\partial w_{2}}{\partial z}\right)\theta(-z-d/2)+w_{2}(-d/2)\delta(z+d/2)\end{array}\label{19}\end{equation}
 is the charge density and $R=\sqrt{(\mathbf{r}-\mathbf{r}')^{2}+(z-z')^{2}}$.
We use the Fourier representations given by equation (\ref{8}) and
the Fourier transform\cite{key-33} \begin{equation}
\frac{e^{i\frac{\omega}{c}\sqrt{r^{2}+z^{2}}}}{\sqrt{r^{2}+z^{2}}}=\sum_{\mathbf{k}}\frac{2\pi i}{\kappa}e^{i\mathbf{kr}}e^{i\kappa\left|z\right|}\,\,\,,\label{20}\end{equation}
where $\kappa=\sqrt{\frac{\omega^{2}}{c^{2}}-k^{2}}$. Then we compute
the electric field from the potentials given by equations (\ref{16})
and (\ref{17}) and use equation (\ref{1}) for $\omega_{0}=0,\,\gamma=0,\,\mathbf{E}_{0}=0$
in order to get integral equations for $\mathbf{v}_{1,2},\, w_{1,2}$.
We define the wavevector $\mathbf{k}_{\perp}$ of magnitude $k$ and
perpendicular to the wavevevctor $\mathbf{k}$, and introduce the
notations $v_{1,2}=\mathbf{k}\mathbf{v}_{1,2}/k$, $v_{1,2}^{\perp}=\mathbf{k}_{\perp}\mathbf{v}_{1,2}/k$.
Doing so, we get the first set of integral equations\begin{equation}
\begin{array}{c}
v_{1}^{\perp}=-\frac{i\omega_{1}^{2}}{2c^{2}\kappa}\int_{d/2}^{\infty}dz'e^{i\kappa\left|z-z'\right|}v_{1}^{\perp}(z')-\frac{i\omega_{2}^{2}}{2c^{2}\kappa}\int_{-\infty}^{-d/2}dz'e^{i\kappa(z-z')}v_{2}^{\perp}(z')\,\,,\,\, z>d/2\,\,\,,\\
\\v_{2}^{\perp}=-\frac{i\omega_{1}^{2}}{2c^{2}\kappa}\int_{d/2}^{\infty}dz'e^{-i\kappa(z-z')}v_{1}^{\perp}(z')-\frac{i\omega_{2}^{2}}{2c^{2}\kappa}\int_{-\infty}^{-d/2}dz'e^{i\kappa\left|z-z'\right|}v_{2}^{\perp}(z')\,\,,\,\, z<-d/2\,\,\,,\end{array}\label{21}\end{equation}
where we dropped out the arguments $\omega,\mathbf{k}$.

Then, from the integral equations for $v_{1,2}$ and $w_{1,2}$ we
notice the relationship \begin{equation}
w_{1,2}=\frac{ik}{\kappa^{2}-\omega_{1,2}^{2}/c^{2}}\frac{\partial v_{1,2}}{\partial z}\,\,\,,\label{22}\end{equation}
which we use to eliminate $w_{1,2}$ from these equations; so, we
are left with the second set of two integral equations in $v_{1,2}$:

for $z>d/2$\begin{equation}
\begin{array}{c}
\frac{c^{2}\kappa^{2}(\omega^{2}-\omega_{1}^{2})}{c^{2}\kappa^{2}-\omega_{1}^{2}}v_{1}=-\frac{i\kappa\omega_{1}^{2}(\omega^{2}-\omega_{1}^{2})}{2(c^{2}\kappa^{2}-\omega_{1}^{2})}\int_{d/2}^{\infty}dz'e^{i\kappa\left|z-z'\right|}v_{1}(z')-\\
\\-\frac{i\kappa\omega_{2}^{2}(\omega^{2}-\omega_{2}^{2})}{2(c^{2}\kappa^{2}-\omega_{2}^{2})}\int_{-\infty}^{-d/2}dz'e^{i\kappa(z-z')}v_{2}(z')+\\
\\+\frac{c^{2}k^{2}\omega_{1}^{2}}{2(c^{2}\kappa^{2}-\omega_{1}^{2})}e^{i\kappa(z-d/2)}v_{1}(d/2)-\frac{c^{2}k^{2}\omega_{2}^{2}}{2(c^{2}\kappa^{2}-\omega_{2}^{2})}e^{i\kappa(z+d/2)}v_{2}(-d/2)\end{array}\label{23}\end{equation}
 and \begin{equation}
\begin{array}{c}
\frac{c^{2}\kappa^{2}(\omega^{2}-\omega_{2}^{2})}{c^{2}\kappa^{2}-\omega_{2}^{2}}v_{2}=-\frac{i\kappa\omega_{1}^{2}(\omega^{2}-\omega_{1}^{2})}{2(c^{2}\kappa^{2}-\omega_{1}^{2})}\int_{d/2}^{\infty}dz'e^{-i\kappa(z-z')}v_{1}(z')-\\
\\-\frac{i\kappa\omega_{2}^{2}(\omega^{2}-\omega_{2}^{2})}{2(c^{2}\kappa^{2}-\omega_{2}^{2})}\int_{-\infty}^{-d/2}dz'e^{i\kappa\left|z-z'\right|}v_{2}(z')-\\
\\-\frac{c^{2}k^{2}\omega_{1}^{2}}{2(c^{2}\kappa^{2}-\omega_{1}^{2})}e^{-i\kappa(z-d/2)}v_{1}(d/2)+\frac{c^{2}k^{2}\omega_{2}^{2}}{2(c^{2}\kappa^{2}-\omega_{2}^{2})}e^{-i\kappa(z+d/2)}v_{2}(-d/2)\end{array}\label{24}\end{equation}
for $z<-d/2$. It is worth observing in deriving these equations the
non-intervertibility of the derivatives and the integrals, according
to the identity \begin{equation}
\frac{\partial}{\partial z}\int_{d/2}^{\infty}dz^{'}f(z^{'})\frac{\partial}{\partial z^{'}}e^{i\kappa\left|z-z^{'}\right|}=\kappa^{2}\int_{d/2}^{\infty}dz^{'}f(z^{'})e^{i\kappa\left|z-z^{'}\right|}-2i\kappa f(z)\label{25}\end{equation}
 for any function $f(z)$, $z>d/2$; a similar identity holds for
$z,z'<-d/2$. It is due to the discontinuity in the derivative of
the function $e^{i\kappa\left|z-z^{'}\right|}$ for $z=z^{'}$. We
can see that these equations become equations (\ref{10}) in the non-retarded
limit by taking formally the limit $c\rightarrow\infty$. However,
this is not so for their dispersion equations, as we shall see below.
One can also see from equations (\ref{21}), (\ref{23}) and (\ref{24})
that the coupling between the two bodies is performed through both
bulk and surface degrees of freedom, in contrast to the non-retarded
situation, where this coupling occurs only through surfaces (equations
(\ref{11})).

We turn now to equations (\ref{21}). Taking the second derivative
with respect to $z$ in these equations we get \begin{equation}
\frac{\partial^{2}v_{1,2}^{\perp}}{\partial z^{2}}+\left(\kappa^{2}-\frac{\omega_{1,2}^{2}}{c^{2}}\right)v_{1,2}^{\perp}=0\,\,\,,\label{26}\end{equation}
which tells that $v_{1,2}^{\perp}$ are a superposition of two waves
$e^{\pm i\kappa_{1,2}z}$, where \begin{equation}
\kappa_{1,2}=\sqrt{\kappa^{2}-\frac{\omega_{1,2}^{2}}{c^{2}}}\,\,.\label{27}\end{equation}
We note that such modes are polaritonic modes, since $\omega^{2}=c^{2}\left(k^{2}+\kappa^{2}\right)=c^{2}\left(k^{2}+\kappa_{1,2}^{2}\right)+\omega_{1,2}^{2}=c^{2}K_{1,2}^{2}+\omega_{1,2}^{2}$,
where $\mathbf{K}_{1,2}=(\mathbf{k},\kappa_{1,2})$, which is the
well-kown dispersion relation for the polaritonic modes. It can also
be written as $\omega^{2}\varepsilon_{1,2}=c^{2}K_{1,2}^{2}$, where
$\varepsilon_{1,2}=1-\omega_{1,2}^{2}/\omega^{2}$ is the dielectric
function for metals. This relation is well-known in the so-called
thery of \char`\"{}effective medium permittivity\char`\"{}. We take
$v_{1,2}^{\perp}=A_{1,2}e^{i\kappa_{1,2}z}$, where $A_{1,2}$ are
amplitudes to be determined. Then, equations (\ref{21}) have non-trivial
solutions for frequencies $\omega$ given by the roots of the dispersion
equation \begin{equation}
e^{2i\kappa d}=\frac{(\kappa_{1}+\kappa)(\kappa_{2}-\kappa)}{(\kappa_{1}-\kappa)(\kappa_{2}+\kappa)}\,\,.\label{28}\end{equation}

Equation (\ref{28}) has a branch of roots for the damped regime (evanescent
modes) $\kappa_{1}=i\alpha_{1}$, $\kappa_{2}=-i\alpha_{2}$, given
by \begin{equation}
\tan\kappa d=\frac{\kappa\left(\alpha_{1}+\alpha_{2}\right)}{\kappa^{2}-\alpha_{1}\alpha_{2}}\,\,\,,\label{29}\end{equation}
where\begin{equation}
\alpha_{1,2}=\sqrt{\frac{\omega_{1,2}^{2}}{c^{2}}-\kappa^{2}}\,\,,\,\,\omega_{1,2}>c\kappa\,\,\,,\label{30}\end{equation}
and $\kappa$ real. Since these modes are damped inside the bodies
and propagating in-between the bodies they may be called surface plasmon-polariton
modes. It is worth noting the correct choice of the sign of the square
root in this case, in order to get the correct behaviour at infinity,
$v_{1}^{\perp}=A_{1}^{\perp}e^{-\alpha_{1}z}$ for $z>d/2$ and $v_{2}^{\perp}=A_{2}^{\perp}e^{\alpha_{2}z}$
for $z<-d/2$. The roots of equation (\ref{29}) can be written as
\begin{equation}
\Omega_{1}=c\sqrt{k^{2}+\frac{\pi^{2}x_{n}^{2}}{d^{2}}}\,\,\,,\label{31}\end{equation}
 where $x_{0}=0$ and $n-1/2<x_{n}<n+1/2$, $n=1,2,3,...$ for $x_{n}<\min\left(\omega_{1},\omega_{2}\right)d/\pi c$.
For identical bodies the roots are given by \begin{equation}
\Omega=c\sqrt{k^{2}+\frac{\pi^{2}n^{2}}{d^{2}}}\label{32}\end{equation}
 for any integer $n=0,1,2...$. They correspond to propagating (polariton)
modes ($\kappa_{1}=\kappa_{2}$ and $\kappa$ all real numbers) and
arise from equation (\ref{28}) for $e^{2i\kappa d}=1$. Equation
(\ref{29}) may have another solution in the vicinity of the vertical
asymptote of the function in its \emph{rhs}, which, however, is irrelevant
for our discussion.

Similarly, $v_{1,2}$ from equations (\ref{23}) and (\ref{24}) obey
the same equation (\ref{26}). We look again for solutions of the
form $v_{1,2}=A_{1,2}e^{i\kappa_{1,2}z}$, where $A_{1,2}$ are amplitudes
to be determined. According to equations (\ref{22}) these modes are
transverse modes, as they should be (for $\kappa_{1,2}$ real). The
relevant dispersion equation is given by \begin{equation}
e^{2i\kappa d}=\frac{(\kappa_{1}+\kappa)(\kappa_{2}-\kappa)(\kappa\kappa_{1}+k^{2})(\kappa\kappa_{2}-k^{2})}{(\kappa_{1}-\kappa)(\kappa_{2}+\kappa)(\kappa\kappa_{1}-k^{2})(\kappa\kappa_{2}+k^{2})}\,\,.\label{33}\end{equation}
We note that this dispersion equation does not become the non-retarded
dispersion equation (\ref{28}) by taking formally the limit $c\rightarrow\infty$. 

An analysis similar to the one performed above for equation (\ref{28})
shows that equation (\ref{33}) has a branch of roots \begin{equation}
\Omega_{2}=c\sqrt{k^{2}+\frac{\pi^{2}y_{n}^{2}}{d^{2}}}\,\,\,,\label{34}\end{equation}
where $y_{0}=0$ and $y_{n}<\min\left(\omega_{1},\omega_{2}\right)d/\pi c$.
They correspond to surface plasmon-polariton modes $\kappa_{1}=i\alpha_{1},\kappa_{2}=-i\alpha_{2}$
and $\kappa$ real. We note that $y_{n}$ may differ from $x_{n}$.
For identical bodies these roots are those given by equation (\ref{32}).
Some other isolated roots may appear, as for instance the one corresponding
to an overall damping, \emph{i.e.} $\kappa_{1}=i\alpha_{1},\,\kappa_{2}=-i\alpha_{2},\,\kappa=i\alpha$,
where $\alpha=\sqrt{k^{2}-\omega^{2}/c^{2}}$, $\omega<ck$. It is
given by \begin{equation}
\Omega_{0}=c\sqrt{k^{2}-\frac{\pi^{2}z_{0}^{2}}{d^{2}}}\,\,\,,\label{35}\end{equation}
 where $\min\left(\omega_{1},\omega_{2}\right)<\pi\sqrt{2}cz_{0}/d<\max\left(\omega_{1},\omega_{2}\right)$.
Such an isolated mode does not contribute significantly to the energy,
so we may neglect it in our subsequent analysis.

We can take the limit $d\rightarrow\infty$ in equation (\ref{33}).
It can be shown that this limit amounts formally to put $e^{2i\kappa d}=0$.\cite{key-23}
We get in this case the surface plasmon-polariton modes corresponding
to a semi-infinite body, given by $\alpha\alpha_{1,2}=k^{2}$, \emph{i.e.}
\begin{equation}
\omega^{2}=\frac{2\omega_{1,2}^{2}c^{2}k^{2}}{\omega_{1,2}^{2}+2c^{2}k^{2}+\sqrt{\omega_{1,2}^{4}+4c^{4}k^{4}}},\label{36}\end{equation}
as derived previously.\cite{key-23} In general, there are problems
with taking formally the limits $d\rightarrow0$ or $d\rightarrow\infty$
in the above equations, as expected. 

It is also worth interesting to look for solutions of the type \begin{equation}
v_{1,2}=A_{1,2}\left[e^{i\kappa_{1,2}z}-e^{\pm i\kappa_{1,2}(d\mp z)}\right]\,\,\label{37}\end{equation}
for equations (\ref{23}) and (\ref{24}), which are vanishing on
the surfaces, $v_{1,2}(\pm d/2)$ (\char`\"{}fixed surfaces\char`\"{}
boundary conditions). In this case, we get again the resonance modes
$\Omega$ given by equation (\ref{32}), irrespective of the bodies
being distinct or identical. In addition, we may get special modes
$\omega=\omega_{1,2}$, $\omega^{2}=c^{2}k^{2}+\omega_{1,2}^{2}$
($\kappa_{1,2}=0$) or $\omega=ck$ ($\kappa=0$), which do not depend
on distance $d$. Other boundary conditions can be put on surfaces
$z=\pm d/2$, and we can get the corresponding eigenmodes.

We note that the dispersion equations (\ref{28}) and (\ref{33})
appear, though in a disguised form, in various formulations of the
fluctuations theory.\cite{key-2},\cite{key-3},\cite{key-5},\cite{key-32}
Within the framework of this theory the dielectric function is included
from the beginning. On the contrary, we recover the dielectric function
in the final results of the present approach, which shows that our
approach is equivalent with the so-called \char`\"{}effective medium
permittivity\char`\"{} theory. 

We pass now to the zero-point energy corresponding to the $\Omega_{1,2}$-eigenmodes
given by equations (\ref{31}) and (\ref{34}), or the $\Omega$-branch
given by equation (\ref{32}) (for identical bodies or \char`\"{}fixed
surfaces\char`\"{}), in the limit $\min\left(\omega_{1},\omega_{2}\right)d/\pi c\gg1$.
These are the only eigenfrequencies which depend on distance $d$.
In the limit $\min\left(\omega_{1},\omega_{2}\right)d/\pi c\gg1$
these modes are dense sets, and it is easy to see that their contributions
to the zero-point energy are equal (corresponding to the two polarizations),
so we can write the total zero-point energy as \begin{equation}
E=\hbar c\sum_{\mathbf{k}n=0}\sqrt{k^{2}+\frac{\pi^{2}x_{n}^{2}}{d^{2}}}\,\,\,,\label{38}\end{equation}
 where $x_{n}$ are defined above; for identical bodies (or for \char`\"{}
fixed surfaces\char`\"{}) $x_{n}=n$. We follow the standard regularization
procedure by removing the ultraviolet divergencies and using the Euler-MacLaurin
formula.\cite{key-34} As it is well-known, the energy thus regularized
reads \begin{equation}
E=\frac{\hbar c}{2\pi}\sum_{k=1}\frac{B_{2k}}{(2k)!}f^{(2k-1)}(x_{0})\,\,\,,\label{39}\end{equation}
where $B_{2k}$ are Bernoulli's numbers and \begin{equation}
f(x)=\int_{0}dk\cdot k\sqrt{k^{2}+\frac{\pi^{2}x^{2}}{d^{2}}}=\frac{1}{2}\int_{\pi^{2}x^{2}/d^{2}}du\cdot\sqrt{u}\,\,.\label{40}\end{equation}
Since $x_{0}=0$ (and $y_{0}=0$), we get the well-known energy $E=-\pi^{2}\hbar cB_{4}/4!d^{3}=-\pi^{2}\hbar c/720d^{3}$
and Casimir force $F=\pi^{2}\hbar c/240d^{4}$ per unit area. The
same result is obtained for the $\Omega$-modes given by equation
(\ref{32}) with $n=0,1,2...$, corrresponding to identical bodies
or the \char`\"{}fixed surfaces\char`\"{} boundary conditions $v_{1,2}(\pm d/2)$.
It is easy to see that for decreasing $\min\left(\omega_{1},\omega_{2}\right)d/\pi c$
the number of $x_{n}$-roots contributing to energy decreases, the
numerical coefficient of the Casimir force decreases gradually, and
the $d^{-4}$-dependence deteriorates, untill a cross-over may occur
to the non-retarded van der Waals-London $d^{-3}$-force. 

The dispersion equations (\ref{28}) and (\ref{33}) hold also for
dielectrics, providing the wavevectors $\kappa_{1,2}$ are changed
according to \begin{equation}
\kappa_{1,2}^{2}\rightarrow\kappa^{2}-\frac{\omega_{1,2}^{2}}{c^{2}}\frac{\omega^{2}}{\omega^{2}-\omega_{01,2}^{2}}\,\,.\label{41}\end{equation}
We can get a usual model of dielectric for $\omega_{01,2}\gg\omega_{1,2}$.
In this case, the wavevectors $\kappa_{1,2}$ become \begin{equation}
\kappa_{1,2}=\sqrt{\kappa^{2}+\frac{\omega_{1,2}^{2}}{\omega_{01,2}^{2}}\frac{\omega^{2}}{c^{2}}}\,\,\,,\label{42}\end{equation}
 and we cannot have anymore surface plasmon-polariton modes (evanescent
modes). In general, under these circumstances, the dispersion equations
(\ref{28}) and (\ref{33}) have no solutions, except for identical
bodies when we may have the $\Omega$-modes given by equation (\ref{32})
($e^{2i\kappa d}=1$) for $n=0,1,2...$. These modes correspond to
propagating polaritons and give again the classical result for the
Casimir force $F=\pi^{2}\hbar c/240d^{4}$ per unit area. Similarly,
for a dielectric-metal pair there is no force, except for boundary
conditions $v_{1,2}(\pm d/2)$ when the resonant $\Omega$-modes given
by equation (\ref{32}) for $n=0,1,2...$ are present. The latter
result holds for any pair of bodies. It is, however, worth stressing
that such results depend on our model of dielectric function for dielectrics,
and, in general, it is necessary to have a quantum-mechanical treatment
for the internal dynamics of the dielectrics.

\section{Discussion and conclusions}

In conclusion, we may say that we have derived here van der Waals-London
and Casimir forces acting between two semi-infinite bodies with parallel
surfaces by calculating the electromagnetic eigenmodes in matter and
estimating their zero-point energy (vacuum fluctuations). We have
adopted well-known, simple, usual models for matter polarization in
metals and dielectrics and made use of the equation of motion for
the polarization in order to get coupled integral equations. The eigenfrequencies
of these equations have been identified and used in calculating the
zero-point energy. In the non-retarded (Coulomb) limit we get the
well-known van der Waals-London $d^{-3}$-force, arising from the
surface plasmons, where $d$ is the distance between the two bodies.
The numerical coefficient of this force acquires various values, depending
on the nature of the bodies and on their being distinct or identical.
When retardation is included we get the Casimir $d^{-4}$-force arising
from surface plasmon-polariton modes (evanescent modes) for a pair
of metals. The classical numerical coefficient of this force ($\pi^{2}/240$)
is obtained for distances much larger than the characteristic wavelengths
($\sim c/\omega_{1,2}$, where $\omega_{1,2}$ are the plasmon frequencies)
of the bodies, and it diminishes gradually for shorter distances,
while the force loses its characteristic $d^{-4}$-dependence. For
a pair of identical dielectrics we get the classical Casimir result
arising from propagating polariton modes. The same result holds for
any pair of bodies with \char`\"{}fixed surfaces\char`\"{} boundary
conditions. 

As it is well-known, the fluctuations theory\cite{key-32} predicts
Casimir forces between any pair of bodies, in contrast with our results,
which give a vanishing force for two distinct dielectrics, for instance.
The difference originates in the circumstance, usually overlooked,
that the equivalent of our dispersion equations (\ref{28}) and (\ref{33})
in the fluctuations theory have no solutions in some cases, as, for
instance, for distinct dielectrics. The usual theorem of meromorphic
functions, applied within the framework of the fluctuations theory,\cite{key-4}-\cite{key-6}
gives then a finite result, but it does not represent the energy of
the eigenmodes. The problem does not appear in the non-retarded regime,
where our results coincide with those of the fluctuations theory.
On the other hand, we must stress again upon the fact that our model
for the dielectric function may not be perfectly adequate for describing
the internal polarization of dielectric matter. Again, this is immaterial
in the non-retarded regime, and we succeeded to compute a $d^{-4}$-van
der Waals-London force between a classical model of polarizable point-like
particle and a semi-infinite body. But our approach fails in this
case in the retarded regime, where a quantum mechanical treatment
is necessary, as in the original attempt in Ref. \cite{key-35}). 

Finally, it is worth noting that the dispersion equations (\ref{28})
and (\ref{33}) can also be obtained by calculating the reflected
field in-between the bodies (fields for semi-infinite bodies).\cite{key-23}
If $r_{1,2}$ are the amplitudes of these fields (for a given polarization),
then the dispersion equations (\ref{28}) and (\ref{33}) are obtained
from $r_{1}=r_{2}e^{2i\kappa d}$. We note that $\left|r_{1,2}\right|^{2}$
are the reflection coefficients, and for two perfectly reflecting
bodies $\left|r_{1}\right|=\left|r_{2}\right|=1$. If we neglect the
phases of the coefficients $r_{1,2}$, and put $r_{1}=r_{2}=1$, we
get the Casimir dispersion equation $e^{2i\kappa d}=1$ ($\Omega$-modes
given by equation (\ref{32})). However, it is precisely these phases
that give the damped surface plasmon-polariton regime, as we have
shown in the present paper, and these phases are not equal in the
damped regime, not even for identical bodies. This is related to the
correct choice of the sign of the square root in $\kappa_{1,2}$,
which, as we have shown here, is $\kappa_{1}=i\alpha_{1}$ and $\kappa_{2}=-i\alpha_{2}$
(equations (\ref{29}) and (\ref{30})). For the propagating regime
(vanishing phases) and identical bodies ($r_{1}=r_{2}$) we get again
the Casimir dispersion equation $e^{2i\kappa d}=1$, as we do for
\char`\"{}fixed surfaces\char`\"{} boundary conditions (in the latter
case irrespective of the bodies).

\textbf{Acknowledgments.} The authors are indebted to the members
of the Theoretical Physics Laboratory at Magurele-Bucharest for valuable
discussions, and to their colleague dr. L. C. Cune for important help
in various stages of this work.

\end{document}